\documentclass[utf8]{frontiersFPHY} 
\setcitestyle{square} 
\usepackage{url,hyperref,lineno,microtype,subcaption}
\usepackage[onehalfspacing]{setspace}
\usepackage{bm,bbm}
\usepackage{mathrsfs}
\usepackage{graphicx}
\usepackage{color}
\usepackage{amsmath}
\usepackage{amssymb,amsfonts}
\usepackage{mathtools}
\usepackage[utf8]{inputenc}
\usepackage{booktabs}
\usepackage{bbm,mathrsfs}

\newcommand{\LL}{\hat{\mathcal{L}}}

\newcommand{\LLB}{\hat{\mathcal{L}}^{\dagger}}
\newcommand{\bbf}{\mathbf{F}(\mathbf{x})}
\newcommand{\bbft}{\mathbf{\tilde{F}}(\mathbf{q})}
\newcommand{\bbftp}{\mathbf{\tilde{F}}(\mathbf{q'})}
\newcommand{\bbfq}{\langle\mathbf{\tilde{F}}(\mathbf{q})\rangle}
\newcommand{\bx}{\mathbf{x}}
\newcommand{\by}{\mathbf{y}}

\newcommand{\sss}{|\mathrm{ss}\rangle}
\newcommand{\sssl}{\langle\mathrm{ss}|}
\newcommand{\lflat}{\langle\text{--}|}
\newcommand{\rflat}{|\text{--}\rangle}

\newcommand{\leftBk}{\langle \psi^L_k|}
\newcommand{\leftKk}{|\psi^L_k\rangle}

\newcommand{\leftKki}{|\psi^L_{k_i}\rangle}

\newcommand{\rightBk}{\langle \psi^R_k|}
\newcommand{\rightKk}{|\psi^R_k\rangle}

\newcommand{\rightKl}{|\psi^R_l\rangle}
\newcommand{\ee}{\mathrm{e}}

\newcommand{\mcp}{\hat{\mathcal{P}}}
\newcommand{\BGamma}{\boldsymbol{\Gamma}}
\newcommand{\bq}{\mathbf{q}}
\newcommand{\eql}{Eq.~(}
\newcommand{\gradient}{\nabla\varphi(\bx)}
\newcommand{\rotation}{\boldsymbol{\vartheta}(\bx)}

\newcommand{\nablaq}{\nabla_{\mathbf{q}}}
\newcommand{\nablaqp}{\nabla_{\mathbf{q}'}}
\newcommand{\bs}{\mathbf{s}}
\newcommand*{\QEDA}{\hfill\ensuremath{\blacksquare}}

\def\keyFont{\fontsize{8}{11}\helveticabold }
\def\firstAuthorLast{Lapolla {et~al.}} 
\def\Authors{Alessio Lapolla\,$^{1}$ and Alja\v{z} Godec\,$^{1,*}$}
\def\Address{$^{1}$Mathematical Biophysics Group, Max Planck Institute
  for Biophysical Chemistry, 37077 G\"ottingen, Germany}
\def\corrAuthor{Alja\v{z} Godec}

\def\corrEmail{agodec@mpibpc.mpg.de}

\begin{document}
\onecolumn
\firstpage{1}

\title[Manifestations of Projection-Induced
  Memory]{Manifestations of Projection-Induced
  Memory: General Theory and the Tilted Single File} 

\author[\firstAuthorLast ]{\Authors} 
\address{} 
\correspondance{} 

\extraAuth{}

\maketitle

\begin{abstract}
Over the years the field of non-Markovian
stochastic processes and anomalous diffusion evolved from a specialized topic to
mainstream theory, which transgressed the realms of physics to
chemistry, biology and ecology. Numerous phenomenological approaches
emerged, which can more or less successfully reproduce or
account for experimental observations in condensed matter, biological
and/or single-particle systems. However, as far as their predictions
are concerned these approaches are not unique, often build on
conceptually orthogonal ideas, and are typically employed on an
\emph{ad hoc} basis.  
It therefore seems timely and desirable to establish a systematic,
mathematically unifying and clean approach starting from more
fine-grained principles. Here we analyze projection-induced ergodic non-Markovian dynamics, both reversible as
well as irreversible, using spectral theory. We investigate dynamical correlations between histories
of projected and latent observables that give rise to memory in
projected dynamics, and rigorously establish conditions under which
projected dynamics is Markovian or renewal. A systematic metric is proposed for quantifying the degree of non-Markovianity.
As a simple, illustrative but non-trivial example we study
single file diffusion in a tilted box, which, for the first time, we solve exactly using the coordinate Bethe
ansatz. Our results provide a solid foundation for a deeper and
more systematic analysis of projection-induced non-Markovian dynamics and anomalous diffusion. 

\tiny
 \keyFont{ \section{Keywords:} Fokker-Planck equation, spectral
   theory, projection operator method, occupation time, single file diffusion, Bethe
   ansatz, free energy landscape} 
\end{abstract}

\section{Introduction}

Over the past decades the field of anomalous diffusion and
non-Markovian dynamics grew to a mainstream physical topic \cite{Ralf1,Ralf2,Sokolov2,Klages,Godec_2014,Ralf3,Franosch,Sokolov,Ralf4,Front}
backed up by a surge of experimental observations
\cite{Woringer,Dix,Krapf,Cox,Ilpo,Rienzo} (the list of works is anything but
exhaustive). From a theoretical point of view the description of
anomalous and non-Markovian phenomena is not universal \cite{Ralf1}
and can be roughly (and judiciously) classified according to the underlying
phenomenology: (i) renewal
continuous-time random walk and fractional Fokker-Planck approaches
\cite{Ralf1,Ralf2,RalfJossi,Sokolov2,Lomholt}, (ii) diffusion in
disordered media \cite{dis1,dis2,dis3,dis4,dis5,dis6,dis7,dis8,dis9},
(iii) generalized Langevin equation descriptions
\cite{GLE,GLE2,GLE3,GLE4,GLE5,GLE6,GLE7,GLE8,GLE9}, (iv) spatially
heterogeneous diffusion \cite{HDP1,HDP2,HDP3,HDP4,HDP5,HDP6,HDP7}, and
more recently also (v) the so-called diffusing diffusivity models
\cite{DD,DD1,DD2,DD3,DD4,DD5,DD6}.    

From a more general first-principles perspective non-Markovian
dynamics in physical systems are always a result of the projection of
nominally deterministic and/or Markovian high-dimensional dynamics to
a lower-dimensional subspace
\cite{Zwanzig1,Zwanzig2,Mori,Grabert77,Haenggi77,Grabert78,Hynes,Haken,Grigolini1,Grigolini2}. The
projection in general induces a dependence of the dynamics on the
initial conditions of the latent degrees of freedom, i.e. those being integrated out, thereby
leading to memory \cite{Zwanzig1,Grabert77,Grabert78,Haenggi77} and
possibly (depending on the system) also to anomalous diffusion
\cite{harris,single,Barkai,Sanders,Olivier1,Olivier2,Lapolla18,Active}.

Hallmarks of
broken Markovianity are the non-validity of the Chapman-Kolmogorov
equation, and on the level of individual trajectories correlations
between histories of projected observables and latent degrees of freedom
\cite{Lapolla18}. The advantage of such an approach is
a deeper understanding and complete control over the origin and nature
of memory effects. The drawback, however, is the inherent difficulty of
integrating out exactly degrees of freedom in a microscopic model,
such that in practice this seems to be only possible for the simplest models, e.g. harmonic
systems (e.g. \cite{Deutch}), comb-models
(e.g. \cite{comb1,comb2,comb3}) or simple obstruction models
\cite{harris,single,Barkai,Sanders,Olivier1,Olivier2,Lapolla18}, to name but a few.     

Here, instead of focusing on the analysis of evolution equations for
projected dynamics \cite{Zwanzig1,Grabert77,Grabert78,Haenggi77} we
focus on the consequences of the projection -- both in a general setting
as well as by means of a simplistic yet non-trivial model of single
file diffusion in a tilted box. Using spectral theory we first present a
rigorous and quite general analysis of the problem and
establish conditions, under which the projection in fact leads to
Markovian or renewal-type
dynamics. We then apply these general results to the analysis of
tagged particle diffusion in a single file confined in a tilted
box. We obtain an exact solution of the full many-body and projected
tagged particle propagators using the
coordinate Bethe ansatz, and provide exact results for tagged particle local time
statistics and correlations between tagged particle
histories. Finally, to asses the degree of non-Markovianity induced
by the projection, we
compute the Kullback-Leibler divergence between the exact tagged particle propagator and the propagator of Markovian
diffusion in the respective free energy landscape, i.e. in the so-called
free energy landscape perspective. Our results provide a deeper
understanding of projection-induced memory and anomalous diffusion and
highlight important pitfalls in applications of free energy landscape-ideas
in absence of a time-scale separation.            

\section{Theory}

\subsection{Notation and Mathematical Preliminaries}
Although all presented result hold identically for discrete-state jump
dynamics governed by a Markovian master equation we will throughout
be interested in projections of continuous (in space as well as time)
Markovian diffusion in a continuous domain $\Omega\in \mathbb{R}^d$ in a vector field $\bbf:\mathbb{R}^d\to \mathbb{R}^d$ (not
necessarily a potential field), which is either nominally confining
(in this case $\Omega$ is open)
or is accompanied by corresponding reflecting boundary conditions at
$\partial \Omega$ (in this case $\Omega$ is closed) thus
guaranteeing the existence of an invariant measure and hence
ergodicity. 
The dynamics are
governed by the (forward) Fokker-Planck operator $\LL:V\to V$ or its
adjoint (or backward) operator $\LLB:W\to W$, where $V$ is a complete normed
linear vector space with elements $f\in C^2(\mathbb{R}^d)$ and $W$ is its dual space. In particular, 
\begin{equation}
  \LL=\nabla\cdot\mathbf{D}\nabla-\nabla\cdot\bbf,\qquad \LLB=\nabla\cdot\mathbf{D}\nabla+\bbf\cdot\nabla,
\label{FPE}
\end{equation}
where $\mathbf{D}$ is the symmetric positive-definite diffusion
matrix. $\LL$ propagates measures $\mu_t(\bx)$ in time, which will throughout
be assumed to posses well-behaved probability density functions
$P(\bx,t)$, i.e. $d\mu_t(\bx)=P(\bx,t)d\bx$ (thereby posing some
restrictions of $\bbf$). Moreover, we assume that $\bbf$ admits the following
decomposition into a potential (irrotational) field $-\mathbf{D}\gradient$
and a non-conservative component
$\rotation$, $\bbf=-\mathbf{D}\gradient+\rotation$ with the two fields being
mutually orthogonal $\gradient\cdot\rotation=0$ \cite{Qian}. By insertion into
\eql\ref{FPE}) one can now easily check that $\LL \ee^{-\varphi(\bx)}=0$,
such that the steady-state solution of the Fokker-Planck equation by
construction does
not depend on the non-conservative part $\rotation$. Before proceeding we first establish the decomposition of the drift field $\bbf$ of the
full dynamics, which with the knowledge
of $\varphi(\bx)$ can be shown to have the form
\begin{equation}
\bbf=-\mathbf{D}\gradient+\ee^{\varphi(\bx)}\mathbf{j}_{\text{ss}}(\bx),
  \label{decomp}
\end{equation}
$\mathbf{j}_{\text{ss}}(\bx)$ denoting the steady-state probability
current and $\rotation\equiv\ee^{\varphi(\bx)}\mathbf{j}_{\text{ss}}(\bx)$ being
incompressible. The proof follows straightforwardly.
We take $\rotation=\bbf+\mathbf{D}\gradient$ and
use $\varphi(\bx)$ to determine the steady-state current
$\mathbf{j}_{\text{ss}}(\bx)=(\rotation-\mathbf{D}\gradient)\ee^{-\varphi(\bx)}+\mathbf{D}\nabla\ee^{-\varphi(\bx)}$,
such that immediately
$\rotation=\ee^{\varphi(\bx)}\mathbf{j}_{\text{ss}}(\bx)$
and in turn follows $\bbf$ in \eql\ref{decomp}). To check for
incompressibility we note that $\mathbf{j}_{\text{ss}}(\bx)$
is by definition divergence free and so
$\nabla\cdot\rotation=\ee^{\varphi(\bx)}(\mathbf{j}_{\text{ss}}(\bx)\cdot\gradient)\equiv\rotation\cdot\gradient=0$,
i.e. $\ee^{\varphi(\bx)}\mathbf{j}_{\text{ss}}(\bx)$ is divergence-free, as claimed.

We define the forward and backward
propagators by $\hat{U}(t)=\mathrm{e}^{\LL t}$ and
$\hat{U}^{\dagger}(t)=\mathrm{e}^{\LLB t}$ such that $\LL$ and $\LLB$
are generators of a semi-group
$\hat{U}(t+t')=\hat{U}(t)\hat{U}(t')$ and
$\hat{U}^{\dagger}(t+t')=\hat{U}^{\dagger}(t)\hat{U}^{\dagger}(t')$, respectively.
For convenience we introduce the bra-ket notation with the 'ket' $|f\rangle$
representing a vector in $V$ (or $W$, respectively) written in
position basis as $f(\bx)\equiv\langle\bx|f \rangle$, and the 'bra'
$\langle g |$ as the integral $\int d\bx g^{\dagger}$. The scalar
product is defined
as $\langle g|f\rangle=\int d\bx g^{\dagger}(\bx)f(\bx)$. Therefore we
have, in operator notation, the following evolution equation for the
conditional probability density function starting from an initial
condition $|p_0\rangle$: $|p_t\rangle=\mathrm{e}^{\LL t}|p_0\rangle$
and, since the process is ergodic, $\lim_{t\to\infty}\mathrm{e}^{\LL
  t}|p_0\rangle=\sss$ defining the equilibrium or non-equilibrium
steady state. In other words, $\LL \sss=0$ and $\sssl\LLB=0$, as a
result of the duality. We also define the (typically non-normalizable)
'flat' state $\rflat$, such that $\langle\bx\rflat=1$ and
$\langle\text{--}|p_t\rangle=1$. Hence,
$\partial_t\langle\text{--}|p_t\rangle=0$ and $\lflat \LL=0$ and
$\LLB\rflat = 0$. We define the Green's function of the process as the
conditional probability density function for a localized initial
condition $\langle\bx|p_0\rangle=\delta(\bx-\bx_0)$ as 
\begin{equation}
G(\bx,t|\bx_0,0) = \langle \bx|\hat{U}(t)|\bx_0\rangle\equiv\langle\bx_0|\hat{U}^{\dagger}(t)|\bx\rangle,
\label{Greens}
\end{equation}
such that the conditional probability density starting from a general
initial condition $|p_0\rangle$ becomes $P(\bx,t|p_0,0)=\langle \bx|\hat{U}(t)|p_0\rangle\equiv\int d\bx_0 p_0(\bx_0) G(\bx,t|\bx_0,0)$.
Moreover, as $\bbf$ is assumed to be sufficiently confining
(i.e. $\lim_{\bx\to\infty}P(\bx,t)=0, \forall t$ sufficiently fast), such that
$\LL$ corresponds to a coercive and densely defined 
operator on $V$ (and $\LLB$ on
$W$, respectively) \cite{Nier,Chupin,Reed}. Finally, $\LL$ is throughout assumed to be \emph{normal},
i.e. $\LLB\LL-\LL\LLB=0$, where for reversible system (i.e. those
obeying detailed balance) we have $\LL\LLB=\LLB\LL=0$. Because any normal compact operator is
diagonalizable \cite{Conway}, we can expand $\LL$ (and $\LLB$) in a
complete bi-orthonormal set of left $\leftBk$ and right $\rightKk$
($\rightBk$ and $\leftKk$, respectively) eigenstates
\begin{equation}
\LL\rightKk=-\lambda_k\rightKk,\qquad \LLB\leftKk=-\alpha_k\leftKk,
\label{spectral}
\end{equation}
with $\text{Re}(\lambda_k)\ge 0$, and according to our definition of the scalar product we have
\begin{equation}
\leftBk\LL\rightKk=-\lambda_k\leftBk\psi^R_k\rangle=\left(\rightBk \LLB \leftKk\right)^{\dagger}=-\alpha_k^{\dagger}\rightBk\psi^L_k\rangle
\label{conjugate}
\end{equation}
and hence the spectra of $\LL$ and $\LLB$ are complex conjugates,
$\alpha_k=\lambda_k^{\dagger}$. Moreover, $\lambda_0=0$,
$|\psi^R_0\rangle=\sss$, $\langle \psi^L_0|=\lflat$, and $\leftBk\psi^R_l\rangle=\delta_{kl}$. Finally, we also have the resolution
of identity $\mathbf{1}=\sum_k\rightKk\leftBk$ and the propagator $\hat{U}(t)=\sum_k\rightKk\leftBk\ee^{-\lambda_kt}$. It follows that the
spectral expansion of the Green's function reads
\begin{equation}
G(\bx,t|\bx_0,0)=\sum_k\psi^R_k(\bx)\psi_k^{L\dagger}(\bx_0)\ee^{-\lambda_kt}\equiv\sum_k\psi^L_k(\bx_0)\psi_k^{R\dagger}(\bx)\ee^{-\lambda_k^{\dagger}t},
\label{SGreen}  
\end{equation}
We now define, $\mcp_{\mathbf{x}}(\BGamma;\bq)$, a (potentially oblique) projection operator into a
subspace of random variables -- 
a mapping
$\bq=\BGamma(\bx):\mathbb{R}^d\to\mathbb{R}^a$ to a subset of coordinates $\bq$ lying in some
orthogonal system in Euclidean space, $\bq\in\Xi(\mathbb{R}^a)\subset\Omega(\mathbb{R}^d)$
with $a< d$. For example, the projection operator applied to some function $R(\bx)\in V$ gives
\begin{equation}
 \mcp_{\bx}(\BGamma;\bq) R(\bx)=\int_{\Omega}d\bx\delta(\BGamma(\bx)-\bq)R(\bx).
  \label{projection}
\end{equation}
The spectral expansion of $\LL$ (and $\LLB$) in the
bi-orthogonal Hilbert space alongside
the projection operator $\mcp_{\bx}(\BGamma;\bq)$ will now allow us to
define and analyze projection-induced non-Markovian dynamics.

\subsection{General Results}

\subsubsection{Non-Markovian Dynamics and (non)Existence of a Semigroup}
Using the projection operator $\mcp_{\bx}(\BGamma;\bq)$
defined in \eql\ref{projection}) we can define the (in general) non-Markovian
Green's function of the projected dynamics as the conditional
probability density of projected dynamics starting from a localized initial
condition $\bq_0$
\begin{equation}
Q_{p_0}(\bq,t|\bq_0,0)=\frac{Q_{p_0}(\bq,t,\bq_0,0)_{p_0}}{Q_{p_0}^0(\bq_0)}\equiv\frac{\mcp_{\bx}(\Gamma;\bq)\mcp_{\bx_0}(\Gamma;\bq_0)G(\bx,t|\bx_0,0)p_0(\bx_0)}{\mcp_{\bx_0}(\Gamma;\bq_0)p_0(\bx_0)},
\label{projected}
\end{equation}
which demonstrates that the time evolution of projected dynamics
starting from a fixed condition $\bq_0$ depends on the initial
preparation of the full system $p_0(\bx_0)$ as denoted by the
subscript. This is a first signature
of the non-Markovian and non-stationary nature of projected dynamics
and was noted upon also in \cite{Haenggi77}. Obviously,
$\int_{\Xi}d\bq Q_{p_0}(\bq,t|\bq_0,0)=1$ for any initial condition
$\bq_0$. We will refer to $\bq$ as the projected degrees of freedom,
whereas those integrated out will be called latent.
For the sake of simplicity we will here mostly limit our discussion to a stationary preparation of the system, i.e.
$p_0(\bx_0)=p_{\text{ss}}(\bx_0)=\langle\bx_0\sss$. In order to avoid
duplicating results we will explicitly carry out the calculation with
the spectral expansion of $\LL$ but note that equivalent results are
obtained using $\LLB$. Using the spectral expansion \eql\ref{SGreen})
and introducing $\Psi_{kl}(\bq)$, the elements of an infinite-dimensional
matrix
\begin{equation}
\Psi_{kl}(\bq)=\leftBk \delta(\BGamma(\bx)-\bq)\rightKl 
\label{element}
\end{equation}
we find from \eql\ref{projected})
\begin{equation}
Q_{p_{\text{ss}}}(\bq,t|\bq_0,0)=\sum_k\Psi_{0k}(\bq)(\Psi_{k0}(\bq_0)/\Psi_{00}(\bq_0))\ee^{-\lambda_kt}
\label{nonMark}
\end{equation}
with $\Psi_{00}(\bq_0)=Q_{p_\text{ss}}^0(\bq_0)$. If one would to
identify $\Psi_{0k}(\bq)=\Psi^R_{0k}(\bq)$ and
$\Psi_{00}(\bq_0)^{-1}\Psi_{0k}(\bq)=\Psi^L_{0k}(\bq)$,
\eql\ref{nonMark}) at first sight looks deceivingly similar to the Markovian Green's
function in \eql\ref{SGreen}).  Moreover, a hallmark of Markovian
dynamics is that it obeys the Chapman-Kolmogorov equation and indeed, since $\leftBk\psi^R_l\rangle=\delta_{kl}$, we find from the spectral expansion
\eql\ref{SGreen}) directly for any $0<t'<t$ that
\begin{equation}
\int_{\Omega} d\bx'
G(\bx,t|\bx',t')G(\bx',t'|\bx_0,0)=\sum_{k,l}\psi^R_k(\bx) \leftBk\psi^R_l\rangle\psi^{L\dagger}_l(\bx_0)\ee^{-\lambda_k(t-t')-\lambda_lt'}\equiv
G(\bx,t|\bx_0,0).
\label{CKE}  
\end{equation}

For non-Markovian dynamics with a stationary $p_0(\bx)$
it is straightforward to prove the following

\noindent \textbf{Proposition 1:} \emph{Let the full system be prepared in a
  steady state, $p_0(\bx)=p_{\mathrm{ss}}(\bx)$, and let non-Markovian Green's function be
defined by \eql\ref{projected}). We take $\Psi_{kl}(\bq)$ as defined in
\eql\ref{element}) and define a scalar product with respect to a
Lebesgue measure $w$ as $\langle f|g\rangle_w\equiv\int d\bx w(\bx)f^{\dagger}(\bx)g(\bx)$. Then Green's function of the projected process
will obey the Chapman-Kolmogorov equation if and only if $\langle
\Psi_{l0}|\Psi_{k0}\rangle_{\Psi_{00}^{-1}}= 0,\forall k,l$.}

We need to prove if and under which conditions 
\begin{equation}
\int_\Xi d\bq'
Q_{p_{\text{ss}}}(\bq,t|\bq',t')Q_{p_{\text{ss}}}(\bq',t'|\bq_0,0)
\label{fail}
\end{equation}
can be equal to $Q_{p_{\text{ss}}}(\bq,t|\bq_0,0)$. As this will
generally not be the case this essentially means that the projected dynamics is in general non-Markovian.
The proof is established by noticing that
$\Psi_{kl}(\bq')=\Psi_{lk}^{\dagger}(\bq')$ such that $\langle \Psi_{l0}|\Psi_{k0}\rangle_{\Psi_{00}^{-1}}\equiv\int_\Xi
d\bq'\Psi_{00}(\bq')^{-1}\Psi_{0l}(\bq')\Psi_{k0}(\bq')$ $\Psi_{00}(\bq)^{-1}d\bq$. As a
result \eql\ref{fail}) can be written
analogously to the first equality in \eql\ref{CKE}) as
\begin{equation}
\sum_{k,l}\Psi_{0k}(\bq)\langle \Psi_{l0}|\Psi_{k0}\rangle_{\Psi_{00}^{-1}}(\Psi_{0l}^{\dagger}(\bq_0)/\Psi_{00}(\bq_0))\ee^{-\lambda_k(t-t')-\lambda_lt'}.
\label{quasi}
\end{equation}
But since the projection mixes all excited eigenstates with $k>0$ (to a
$k$-dependent extent) with the left
and right ground states (see \eql\ref{element})), the orthogonality
between $\Psi_{00}(\bq)^{-1/2}\Psi_{0l}(\bq)$ and
$\Psi_{00}(\bq)^{-1/2}\Psi_{k0}(\bq)$ is in general lost, and $\langle
\Psi_{l0}|\Psi_{k0}\rangle_{\Psi_{00}^{-1}}\ne 0$ for $k\ne
l$ as claimed above. The Chapman-Kolmogorov equation can hence be
satisfied if  and only if $\langle\Psi_{l0}|\Psi_{k0}\rangle_{\Psi_{00}^{-1}}=0$ for all
$k\ne l$. \QEDA

However, even  if
$\langle\Psi_{l0}|\Psi_{k0}\rangle_{\Psi_{00}^{-1}}=0,\forall k \ne l$ this does not guarantee that the projected process is actually Markovian (see
\cite{Feller,Haenggi_B} for observations made in the consideration of
specific model examples). The computation of higher-order probability
densities is necessary in order to check for Markovianity.  

\subsubsection{When is the Projected Dynamics Markovian or Renewal?}

\emph{A) Projected Dynamics is Markovian}

A particularly useful aspect of the present spectral-theoretic approach is its ability to
establish rigorous conditions for the emergence of (exactly) Markovian and
(exactly) renewal-type dynamics from a microscopic, first principles point of
view. Note that in this section we assume a general, non-stationary preparation of the system
(i.e. $p_0(\bx_0)\ne p_{\text{ss}}(\bx_0)$). By inspection of Eqs.~(\ref{element}) and
(\ref{nonMark}) one can establish that:

\noindent\textbf{Theorem 2:} \emph{The necessary and
sufficient condition for the projected dynamics to be Markovian if is
that the projection $\mcp_{\mathbf{\bx}}(\BGamma;\bq)$ (whatever its form) nominally
projects into the nullspace of latent dynamics. In other words,
the latent and projected dynamics remain decoupled and orthogonal
for all times. This means that
(i) there exists a bijective map $\by=f(\bx)$ 
to a decomposable coordinate system $\by=(\bq,\bq'')$, in which the forward generator decomposes to
$\LL=\LL_p+\LL_l$, where $\LL_p$ only acts and depends on the
projected degrees of freedom
$\bq\in\Xi(\mathbb{R}^a)\subset\Omega(\mathbb{R}^d)$ with $a<d$
and $\LL_l$ only acts and depends on the latent coordinates
$\bq''\in\Xi^c(\mathbb{R}^d)\subset\Omega(\mathbb{R}^d)$ (with
, $\Xi\cap\Xi''=\emptyset$, $\Omega=\Xi\cup\Xi''$), (ii) the boundary
conditions on $\partial \Xi$ and $\partial \Xi^c$ are decoupled, and (iii) the projection
operator $\mcp_{\mathbf{\by}}(\cdot;\bq)=\int d\bq''$ onto the subset of coordinates
$\bq\in\Xi(\mathbb{R}^{a})\subset\Omega$ corresponds to an integral over the subset of latent
coordinates $\bq''\in\Xi^c(\mathbb{R}^{d-a})\subset\Omega$, which does
not mix projected and latent degrees of freedom, or alternatively $\LL_l p_0(\bq_0,\bq_0'')=0$.}

The proof is rather straightforward and follows from the fact that if (and only
if) the projected dynamics is
Markovian it must be governed as well by a formal (Markovian) Fokker-Planck
generator $\LL_p$ as in \eql\ref{FPE}), in which the projected and latent degrees of freedom are
separable $\LL=\LL_p+\LL_l$, and that the full Hilbert space is a direct
sum of Hilbert spaces of the $V=V_p\oplus V_l$, that is $\LL:V \to V$,
$\LL_p:V_p\to V_p$ and $\LL_l:V_l\to V_l$ and $V_p\cap V_l=\emptyset$. This also requires that
there is no boundary condition coupling vectors from $V_p$
and $V_l$. In turn this implies assertion (i) above. If
$\mcp_{\by}(\cdot;\bq)$ is such that it does not mix eigenfunctions
in $V_p$ and $V_l$ (i.e. it only involves vectors from $V_p$) then
because of bi-orthonormality and the fact that $\lflat \LL=0$ the
projected Green's function in full space $Q(\bq,t|\bq_0)$ for
$\bq\in\Xi(\mathbb{R}^a)$ will be identical to the full Green's
function in the isolated domain $G(\bx,t|\bx_0)$ for
$\bx\in\Xi(\mathbb{R}^a)$ and the non-mixing
condition is satisfied.  The effect is the same if the latent degrees of freedom already start
in a steady state, $\LL_l p_0(\bq_0,\bq_0'')=0$. This establishes
sufficiency. However, as soon as the projection mixes the two Hilbert spaces $V_p$
and $V_l$, the generator of projected dynamics will pick up
contributions from $\LL_l$ and will, upon integrating out the latent
degrees of freedom, not be
Markovian. This completes the proof.\QEDA

\emph{B) Projected Dynamics is Renewal}

We can also rigorously establish sufficient conditions for the projected dynamics
to poses the renewal property. Namely, the physical notion of a
waiting time or a random change of time-scale (see,
e.g. \cite{Ralf2,Sokolov2}) can as well be attributed a microscopic
origin. The idea of a random waiting time (or a random change of time
scale) nominally implies a
period of time and thereby the existence of some sub-domain, during which
and within the latent degrees evolve while the projected dynamics does not change.
For this to be the case the latent degrees of freedom must be perfectly
orthogonal to the projected degrees of freedom, both in the two
domains as well as on their boundaries (a prominent simple example is
the so-called comb model \cite{comb1,comb2,comb3}). Moreover, the projected
degrees of freedom evolve only when the latent degrees of freedom
reside in some subdomain $\Upsilon\subset\Xi^c(\mathbb{R}^{d-a})$. In turn, this means
that the dynamics until a time $t$ ideally partitions between projected and latent degrees of
freedom, which are coupled solely by the fact that the total time
spent in each must add to $t$, which effects the waiting time. In a comb-setting 
the motion along the backbone occurs only when the
particle is in the center of the orthogonal plane. In the context of a
low-dimensional projection of ergodic Markovian dynamics, we can in fact
prove the following general theorem:

\noindent\textbf{Theorem 3:} \emph{Let there exists a bijective map $\by=f(\bx)$  to a decomposable
  coordinate system $\by=(\bq,\bq'')$ as in A) with the projected $\bq\in\Xi(\mathbb{R}^a)$ and latent degrees of freedom 
  $\bq''\in\Xi^c(\mathbb{R}^{d-a})\equiv\Omega(\mathbb{R}^d)\setminus\Xi(\mathbb{R}^a)$. 
 Furthermore, let $\Upsilon\subset\Xi^c(\mathbb{R}^{d-a})$ and 
                                let $\mathbbm{1}_{\Upsilon}(\bq'')$ denote the indicator function of
the region $\Upsilon$ (i.e. $\mathbbm{1}_{\Upsilon}(\bq'')=1$ if
$\bq''\in\Upsilon$ and zero otherwise). Moreover, let the full system
be prepared in an initial condition $p_0(\bq,\bq'')$.
Then a sufficient condition for renewal-type dynamics is (i) that the
forward generator in $(\bq,\bq'')$ decomposes $\LL= \mathbbm{1}_{\Upsilon}(\bq'')\LL_p+\LL_l$, and where $\LL_p$ only acts and depends on $\bq$
and $\LL_l$ only acts and depends on $\bq''$, and (ii) the boundary conditions do not
cause a coupling of latent and projected degrees of freedom (as in the
Markov case above).} 

The proof can be established by an explicit
construction of the exact evolution equation for the projected variables. 
Let $G_l(\bq'',t|\bq''_0)$ denote the
Green's functions of the Markovian problem for the latent degrees of freedom,
$G_l(\bq'',t|\bq''_0)=\langle\bq''|\ee^{\LL_lt}|\bq''_0\rangle=\sum_k\langle\bq''|\psi^{l,R}_k\rangle\langle
\psi^{l,L}_k|\bq_0''\rangle\ee^{-\lambda_k^lt}$ and let
$\tilde{g}(s)=\int_0^\infty\ee^{-st}g(t)dt$ denoted the Laplace
transform of a function $g(t)$. The projection operator in this case
corresponds to $\mcp_{\bq''}(\cdot;\bq)=\int_{\Xi^c}d\bq''$. We
introduce the shorthand notation
$\overline{p}_0(\bq)=\int_{\Xi^c}d\bq_0''p_0(\bq_0,\bq_0'')$ and
define the conditional initial probability density $p_0(\bq_0''|\bq_0)=p_0(\bq_0,\bq_0'')/\overline{p}_0(\bq_0)$.
The Green's function of projected dynamics becomes
$Q_{p_0}(\bq,t|\bq_0)=\int_{\Xi^c}d\bq''\int_{\Xi^c}d\bq_0''G(\bq,\bq'',t|\bq_0,\bq''_0)p_0(\bq_0,\bq''_0)/\overline{p}_0(\bq_0)$. We
then have the following lemma:

\noindent\textbf{Lemma 4:} \emph{Under the specified assumptions $Q(\bq,t|\bq_0)$ exactly obeys the renewal-type non-Markovian Fokker-Planck
equation}  
\begin{equation}
\partial_tQ_{p_0}(\bq,t|\bq_0)= \int_0^{t}d\tau K_{p_0}(t-\tau)\LL_pQ_{p_0}(\bq,\tau|\bq_0),
\label{FPEnm}
\end{equation}
\emph{with the memory kernel
 \begin{eqnarray} 
  K_{p_0}(t)&=&(\delta(t) +
  \partial_t)\int_{\Upsilon}d\bq''\int_{\Xi^c}d\bq''_0p_0(\bq_0''|\bq_0)\langle\bq''|\ee^{\LL_lt}|\bq''_0\rangle\nonumber\\
  &=&\sum_k\left(\int_{\Xi^c}d\bq_0''\psi^{l,L
  \dagger}_k(\bq''_0)p_0(\bq_0''|\bq_0)\right)\left(\int_{\Upsilon}d\bq''\psi^{l,R}_k(\bq'')\right)(\delta(t)-\lambda_k^l\ee^{-\lambda_k^lt})
\label{memory}
\end{eqnarray}
that is independent of
$\bq$ . Moreover, $Q(\bq,t|\bq_0)>0$
for all $t>0$ and for all $\bq,\bq_0\in\Xi$.}

To prove the lemma we Laplace transform equation ($t\to u$)
$\partial_tG(\bq,\bq'',t|\bq_0,\bq''_0)=\LL
G(\bq,\bq'',t|\bq_0,\bq''_0)$ and realize that the structure of $\LL$
implies that its solution with initial condition
$\delta(\bq-\bq_0)\delta(\bq''-\bq''_0)$ in Laplace space factorizes 
$\tilde{G}(\bq,\bq'',u|\bq_0,\bq''_0)=f_u(\bq|\bq_0)g_u(\bq''|\bq_0'')$
with $g_u$ and $f_u$ to be determined. Note that $\int_{\Xi} d\bq\int_{\Xi^c} d\bq''\tilde{G}(\bq,\bq'',u|\bq_0,\bq''_0)=\int_{\Xi} d\bq
f_u(\bq|\bq_0)\int_{\Xi^c}d\bq''g_u(\bq''|\bq_0'')=u^{-1}$ and we can
chose, without any loss of generality that $\int_{\Xi} d\bq
f_u(\bq|\bq_0)=1$. Plugging in the factorized ansatz
and rearranging leads to
\begin{equation}
g_u(\bq''|\bq''_0)\left(u f_u(\bq|\bq_0)- \mathbbm{1}_{\Upsilon}(\bq'')\LL_pf_u(\bq|\bq_0)\right)-f_u(\bq|\bq_0)\LL_lg_u(\bq''|\bq''_0)-\delta(\bq-\bq_0)\delta(\bq''-\bq_0'')=0.
\label{auxe}  
\end{equation}
Noticing that $\int_\Xi d\bq \LL_pf(\bq|\bq_0)=0$ as a result of the divergence
theorem (as we assumed that $\bbf$ is strongly confining implying that
the current vanishes at the boundaries) 
we obtain, upon
integrating \eql\ref{auxe}) over $\bq$
\begin{equation}
ug_u(\bq''|\bq''_0)-\delta(\bq''-\bq_0'')-\LL_lg_u(\bq''|\bq''_0)=0,
\label{auxe2}  
\end{equation}
implying that $g_u(\bq''|\bq_0'')=\tilde{G}_l(\bq'',u|\bq_0'')$. As
$\tilde{G}_l(\bq'',u|\bq_0'')$ is the Laplace image of a Markovian Green's function
we use $\int_{\Xi^c}d\bq''\tilde{G}_l(\bq'',u|\bq_0'')=u^{-1}$ in order to
deduce that $\tilde{Q}_{p_0}(\bq,u|\bq_0)=f_u(\bq|\bq_0)/u$.
The final step involves using the identified functions $f_u$ and $g_u$
in \eql\ref{auxe}), multiplying with
$p_0(\bq_0''|\bq_0)$, integrating over
$\bq''$ and $\bq_0''$ while using the divergence
theorem implying
$\int_{\Xi^c}d\bq''\LL_l\tilde{G}_l(\bq'',u|\bq_0'')=0$ (as before)
 to obtain
 \begin{equation}
u\tilde{Q}_{p_0}(\bq,u|\bq_0)-\delta(\bq-\bq_0)=\left(u\int_{\Upsilon}d\bq''\int_{\Xi^c}d\bq_0''\tilde{G}_l(\bq'',u|\bq_0'')p_0(\bq_0''|\bq_0)\right)\LL_p\tilde{Q}_{p_0}(\bq,u|\bq_0).
  \label{master}
 \end{equation}
Finally, since the Laplace transform of
$\partial_tg(t)+\delta(t)g(0)$ corresponds to $u\tilde{g}(u)$, 
taking the inverse Laplace
transform of \eql\ref{master}) finally leads to Eqs.~(\ref{FPEnm}) and
\ref{memory}) and completes
the proof of the lemma, since now we can take $Q_{p_0}(\bq,t|\bq_0)>0$ by
definition because \eql\ref{FPEnm}) is an identity of \eql\ref{FPE})
integrated over $\bq''$. Moreover, the rate of change of the Green's function
$Q_{p_0}(\bq,t|\bq_0)$ in \eql\ref{FPEnm}) depends, at any instance $t$,
position $\bq$ and for any initial condition $\bq_0$ only on the
current position $\bq$ and a waiting time (or random time-change)
encoded in the memory kernel $K(t)$; $Q_{p_0}(\bq,t|\bq_0)$ is the Green's
function of a renewal process. This completes the proof of sufficiency.\QEDA

Furthermore, for the situation where the full system is prepared in a
stationary state, i.e. $p_0(\bx)=p_{\mathrm{s}}(\bx)$, we have
the following corollary:

\noindent \textbf{Corollary 5:} \emph{Let the system and projection be defined as in Theorem 3. If the
  full system is prepared such that the latent degrees of freedom are
  in a stationary state
  $p_0(\bq_0,\bq''_0)$, such that  $\LL_lp_0(\bq''_0|\bq_0)=0, \forall
  \bq_0\in\Xi$ and hence also
  $\overline{p}_{0}(\bq_0'')=\overline{p}_{\mathrm{ss}}(\bq_0'')$,
  then $p_0(\bq''_0|\bq_0)=\psi^{l,R}_0(\bq''_0)$ and consequently
  $K_{p_0}=\delta(t)\int_{\Upsilon}d\bq_0''\overline{p}_{\mathrm{ss}}(\bq_0'')$,
  and therefore the projected dynamics is Markovian. Moreover, if the system is prepared such that the latent degrees of freedom are
  not in a stationary state, i.e.
  $p_0(\bq_0|\bq''_0)\ne\overline{p}_{\mathrm{ss}}(\bq_0''), \forall \bq_0$, there
  exists a finite time $t_M>0$ after which the dynamics will
  be arbitrarily close to being Markovian.}

The proof of the first part follows from the bi-orthogonality of eigenfunctions of
latent dynamics
$\langle\psi^{l,R}_k|\psi^{l,R}_0\rangle=\delta_{k,0}$, rendering all
terms in \eql\ref{memory}) in Lemma 4 identically zero except for $k=0$ with
$\lambda_k^l=0$. The second part is established by the fact that for
times $t_M\gg 1/\lambda_1^l$, with $\lambda_1^l$ being the largest
(i.e. least negative) non-zero eigenvalue, all terms but the $k=0$
term in \eql\ref{memory}) in Lemma 4 become arbitrarily small. \QEDA 

\begin{figure}[!!ht]
\begin{center}  
\includegraphics[width=10.5cm]{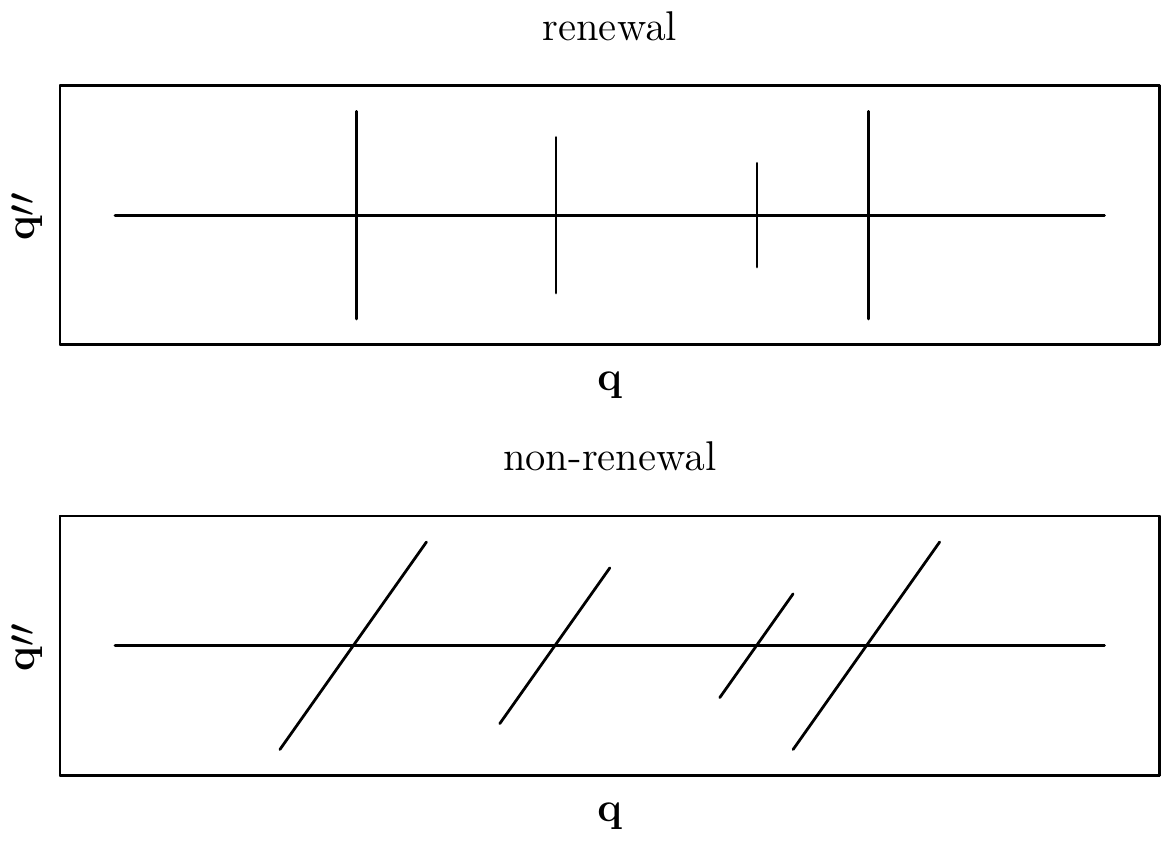}
\caption{Schematics of a generalized comb model. For the sake of clarity only a
  couple of side-branches are shown, whereas the model is to be
  understood in the sense of densely populated side-branches. (top) As long as the projected
  $q$ and latent $q''$ degree of freedom remain orthogonal, the
  projected dynamics will be of renewal-type. However, as soon as this
ceases to be the case the projected dynamics will not be renewal.}
\label{fig:0}
\end{center}
\end{figure}

Having established sufficiency, we now also comment on necessity of
the conditions (i) and (ii) above for renewal dynamics. It is
clear that the splitting of $\LL$ into $\LL_p$ and
$\LL_l$, where $\LL_l$ does not act nor depend on projected
variables, is also necessary condition for
renewal. This can be established by contradiction as loosening
these assumptions leads to dynamics that is not renewal. This can be
understood intuitively,
because it must hold that the
latent degrees of freedom remain entirely decoupled from the projected
ones (but not vice versa) and that the motion along both is mutually orthogonal. To
illustrate this think of the paradigmatic comb model (see schematic in Fig.\ref{fig:0})
\cite{comb1,comb2,comb3} and realize that renewal will be violated as
soon as we tilt the side-branches for some angle from being orthogonal
to the backbone. 

However, since it is difficult to establish the
most general class of admissible functions $h(\bq'')$ used in
$\LL=h(\bq'')\LL_p+\LL_l$, 
we are not able to prove necessity. Based on the present analysis it
seems somewhat difficult to systematically relax the assumptions for
projected dynamics to be renewal without assuming, in addition, some
sort of spatial discretization. We therefore hypothesize that the
sufficient conditions stated in Theorem 3, potentially with some
additional assumptions on $h(\bq'')$ are also necessary conditions.  Notably, however, microscopic
derivations of non-Markovian master equations of the form given in
\eql\ref{FPEnm}) often start in discretized space or \emph{ad
  hoc} introduce a random change in time scale (see e.g. \cite{Ralf2,RalfJossi,Sokolov_master}).

\subsubsection{Markovian Approximation and the Degree of non-Markovianity}
In order to quantify the degree of non-Markovianity induced by the
projection we propose to compare the full non-Markovian dynamics with
projected dynamics evolving under a complete time-scale separation,
i.e. under the assumption of all latent degrees of freedom being in
the stationary state. To do so we proceed as follows.
The projected coordinates $\bq$ are now assumed to represent a subset of another
$d$-dimensional orthogonal system in Euclidean space $\mathbf{q'}\in\mathbb{R}^d$, and we
assume the map $\bq'(\bx)$ is bijective. We denote the
conditional probability density in this system by
$G'(\bq',t|\bq'_0,0)$. The underlying physical idea is that an observer can only see the projected dynamics, which
since it is non-Markovian stems from a projection but not
necessarily onto Cartesian coordinates. Therefore, from a physical
perspective not too much
generality seems to be lost with this assumption. 

As a concrete example one can consider the
non-spherically symmetric Fokker-Planck process in a sphere,
corresponding to the full Markovian parent system projected
onto angular variables (either one or both). This way one
first transforms from $\bx\in\mathbb{R}^3$ to spherical coordinates
$\bq'=(r,\phi,\theta)$ and then, e.g. projects on the the lines $\bq=\phi\in[0,2\pi)$. 

Since the transformation of the Fokker-Planck equation under a general
change of coordinates is well-known \cite{Risken} the task is actually
simple. Under the complete map $\bq'=\BGamma(\bx)$ with
$\BGamma:\mathbb{R}^d\to\mathbb{R}^d$ the forward Fokker-Planck
operator in
\eql\ref{FPE}) transforms as
$\LL'=\nablaqp\otimes\nablaqp:\mathbf{\tilde{D}}(\bq')-\nablaqp\cdot\bbftp$,
where $\otimes$ and $:$ denote, respectively, the tensor and double-dot product, and the transformed drift
field and diffusion tensor can be written as
\begin{equation}
(\bbftp)_k=\sum_{i=1}^d\frac{\partial q'_k}{\partial
    x_i}\mathbf{F}_i+\sum_{i,j=1}^dD_{ij}\frac{\partial^2
    q_k}{\partial x_i\partial x_j},\quad
(\mathbf{\tilde{D}}(\bq'))_{kl}=\sum_{i,j=1}^d D_{ij}\frac{\partial q'_k}{\partial
    x_i}\frac{\partial q'_l}{\partial
    x_j}.
  \label{FPEt}
\end{equation}
We note that unless the mapping is linear, the old diffusion
  matrix affects the new drift vector and the diffusion matrix picks up a
  spatial dependence. For an excellent account of the transformation
  properties in the more general case of a position
  dependent diffusion matrix (i.e. $\mathbf{D}\to\mathbf{D}(\bx)$)
  we refer the reader to  \cite{Polettini}. 
We now want to marginalize over the remaining (i.e. non-projected) coordinates
$\bq''\in\Omega\setminus\Xi$ but beforehand make the Markovian
approximation 
$G'(\bq',t|\bq_0,0)\approx Q_M(\bq,t|\bq_0)p_{\text{ss}}(\bq'')$.
Then we have $\LL'G'(\bq',t|\bq_0,0)\approx
p_{\text{ss}}(\bq'')\LL'Q_M(\bq,t|\bq_0)$, implying that the
operator $\LL'$ approximately splits into one part operating on the projected
coordinates alone, $\LL'_M$, and one operating only on the latent stationary
coordinates, $\LL''$, for which $\LL''p_{\text{ss}}(\bq'')=0$. The
physical idea behind the Markovian approximation is that the latent
degrees of freedom relax infinitely fast compared to the projected
ones. Therefore, we
can straightforwardly average the Fokker-Planck operator over the stationary latent
coordinates $\bq''$, $\langle\LL_M'\rangle_{\bq''}$, where we have
defined the latent averaging operation $\langle\cdot\rangle_{\bq''}\equiv\int
d\bq''p_{\text{ss}}(\bq'')\cdot$. Note that the remaining dependence of
$\LL'$ on the latent stationary coordinates $\bq''$ is only due to
$\bbftp$ and $\mathbf{\tilde{D}}(\bq')$. The averaged drift field and
diffusion matrix now become
\begin{equation}
\bbfq_k=\sum_{i=1}^d\left\langle\frac{\partial q'_k}{\partial
    x_i}\mathbf{F}_i\right\rangle_{\bq''}+\sum_{i,j=1}^dD_{ij}\left\langle\frac{\partial^2
    q_k}{\partial x_i\partial x_j}\right\rangle_{\bq''},\quad
\langle\mathbf{\tilde{D}}(\bq)\rangle_{kl}=\sum_{i,j=1}^d D_{ij}\left\langle\frac{\partial q'_k}{\partial
    x_i}\frac{\partial q'_l}{\partial
    x_j}\right\rangle_{\bq''}.
  \label{FPEta}
\end{equation}
We can further decompose the effective drift field into a
conservative and a non-conservative part
\begin{equation}
\left\langle\frac{\partial q'_k}{\partial
    x_i}\mathbf{F}_i\right\rangle_{\bq''}=-\left\langle\frac{\partial q'_k}{\partial
    x_i}\left(\mathbf{D}\nabla\varphi\right)_i\right\rangle_{\bq''}+\left\langle\ee^{\varphi}\frac{\partial q'_k}{\partial
    x_i}(\mathbf{j}_{\text{ss}})_i\right\rangle_{\bq''},
  \label{split}
\end{equation}
which establishes the Markovian approximation also for a broad class
of irreversible systems.
The approximate effective Fokker-Planck operator for the projected
dynamics in turn reads
\begin{equation}
\langle\LL'\rangle_{\bq''}=\nablaq\otimes\nablaq:\langle\mathbf{\tilde{D}}(\bq)\rangle_{\bq''}-\nablaq\cdot\langle\bbft\rangle_{\bq''}.
\label{FPEp}
\end{equation}
By design the kernel of $\langle\LL'\rangle_{\bq''}$ is equal to
$p_{\text{ss}}(\bq)\equiv\mcp_{\mathbf{x}}(\BGamma;\bq)p_{\text{ss}}(\bx)$, hence
$\langle\LL'\rangle_{\bq''}$ governs the relaxation towards the
steady-state density (not necessarily equilibrium) evolving from some initial
state $\bq_0$ in the Markovian approximation with the corresponding
Green's function
$Q_M(\bq,t|\bq_0,0)\equiv\langle\bq|\ee^{\langle\LL'\rangle_{\bq''}
  t}|\bq_0\rangle$.

In order to quantify the departure of the exact dynamics from the
corresponding Markovian behavior we propose to evaluate the
Kullback-Leibler divergence between the Green's functions of the exact
and Markovian propagator as a function of time
\begin{equation}
\mathcal{D}_t(Q||Q_M)=\int_{\Xi}d\bq Q(\bq,t|\bq_0,0)\ln\left(\frac{Q(\bq,t|\bq_0,0)}{Q_M(\bq,t|\bq_0,0)}\right).
\label{KL}
\end{equation}  
 By definition $\mathcal{D}_t(Q||Q_M)\ge 0$ and since the non-Markovian behavior of the exact projected dynamics is
 transient with a life-time $\lambda_1^{-1}$, we have that
 $\lim_{t\to\infty}\mathcal{D}_t(Q||Q_M)=0$. Our choice of quantifying the departure of the exact dynamics from the
corresponding Markovian behavior is not unique. The
Kullback-Leibler divergence introduced here can hence be used to
quantify how fast the correlation of the latent degrees of freedom
with the projected
degrees of freedom dies out. Notably, in a related manner the Kullback-Leibler divergence was also
used in the context of stochastic thermodynamics in order to disprove
the hypothesis about the monotonicity of the entropy production as a
general time evolution principle \cite{Polettini2}.

\subsubsection{Functionals of Projected Dynamics}
In order to gain deeper insight into the origin and manifestation of
non-Markovian behavior it is instructive to focus on the statistics of
time-average observables, that is functionals of
projected dynamics. As in the previous sections we assume that the
full system was prepared in a (potentially non-equilibrium current-carrying) steady state.
To that end we have, using Feynman-Kac theory, recently proven a theorem connecting any bounded additive
functional $\Phi_t[\bq(\tau)]=t^{-1}\int_0^tZ(\bq(\tau))d\tau$  (with
a function  $Z:\Xi(\mathbb{R}^a)\to\mathbb{R}$ locally strictly bounded in $\Xi$) of projected dynamics $\bq(\tau)$ of a parent
Markovian diffusion $\bx(t)$ to the eigenspectrum of the Markov
generator of the full dynamics $\LL$ or $\LLB$ \cite{Lapolla18}. The
central quantity of the theory is $\theta_t(\bs)$, the so-called local time fraction
spent by a trajectory $\bq(\tau)$ in a infinitesimal volume element $d\bs$
centered at $\bs$ up until a time $t$ enabling 
\begin{equation}
\theta_t(\bs)=t^{-1}\int_0^td\tau\mathbbm{1}_{\bs}(\bq(\tau))\quad\to\quad
\Phi_t[\bq(\tau)]=\int_\Xi d\bs Z(\bs)\theta_t(\bs),
\label{localt}  
\end{equation}
where the indicator function $\mathbbm{1}_\bs(\bq)=1$ if $\bq=\bs$ and
zero otherwise. We
are here interested in the fluctuations of $\theta_t(\bs)$ and correlation
functions between the local time fraction of a projected observable
$\bq(t)$ at a point $\bs$ and $\theta''(\bs')$, the local time some latent (hidden) observable
$\bq''(t)$ a the point $\bs'$:
\begin{equation}
\sigma^2_t(\bs)=\langle\theta^2_t(\bs)\rangle-\langle\theta_t(\bs)\rangle^2,\quad \mathcal{C}_t(\bs;\bs')=\langle\theta_t(\bs)\theta''_t(\bs')\rangle-\langle\theta_t(\bs)\rangle\langle\theta''_t(\bs')\rangle,
\label{varcor}  
\end{equation}
where $\langle\cdot\rangle$ now denotes the average over all forward paths
starting from the steady state $|\bq_0\rangle=\sss$ (and ending anywhere,
i.e. $\langle \bq|=\lflat$), or, using the backward approach, all
paths starting in the flat state $|\bq\rangle=\rflat$ and propagating backward in
time towards the steady state $\langle \bq_0|=\sssl$. 
We note that any correlation function of a general additive bounded
functional $\Phi^i_t[\bq(\tau)]$ of the form
$\langle\Phi^i_t[\bq(\tau)]\Phi^j_t[\bq''(\tau)]\rangle$ (as well as
the second moment of $\Phi^i_t[\bq(\tau)]$)  follows directly from the
local time fraction, namely,
$\langle\Phi^i_t[\bq(\tau)]\Phi^j_t[\bq''(\tau)]\rangle=\int_\Xi\int_\Xi
d\bs
d\bs'Z_i(\bs)Z_j(\bs')\langle\theta_t(\bs)\theta''_t(\bs')\rangle$. For details of the theory and corresponding proofs
please see \cite{Lapolla18}, here we will simply state the main theorem: 

\emph{Let the Green's function of the full parent dynamics $\bx(t)$ be given
by \eql\ref{SGreen}) and the local time fraction $\theta_t(\bs)$ by
\eql\ref{localt}), then the variance and and correlation function
defined in \eql\ref{varcor}) is given exactly as}
\begin{eqnarray}
  \sigma^2_t(\bs)&=& 2\sum_{k>0}\frac{\lflat\mathbbm{1}_{\bs}\rightKk\leftBk\mathbbm{1}_{\bs}\sss}{\lambda_kt}\left(1-\frac{1-\ee^{-\lambda_kt}}{\lambda_kt}\right)\nonumber\\
 \mathcal{C}_t(\bs;\bs')&=&\sum_{k>0}\frac{\lflat\mathbbm{1}_{\bs}\rightKk\leftBk\mathbbm{1}''_{\bs'}\sss+\lflat\mathbbm{1}''_{\bs'}\rightKk\leftBk\mathbbm{1}_{\bs}\sss}{\lambda_kt}\left(1-\frac{1-\ee^{-\lambda_kt}}{\lambda_kt}\right),
\label{local_spectral}  
\end{eqnarray}
\emph{and analogous equations are obtained using the backward approach
\cite{Lapolla18}.}

The usefulness of \eql\ref{local_spectral}) can be understood as follows. By varying $\bs$ and $\bs'$ one can establish \emph{directly} the regions
in space responsible for the build-up (and subsequent decay) of memory
in projected dynamics and simultaneously monitor the fluctuations of
the time spent of a projected trajectory in said
regions. Note that because the full process is assumed to be ergodic,
the statistics of $\theta_t(\bs)$ will be asymptotically Gaussian
obeying the large deviation principle.
This concludes our general results. In
the following section we apply the theoretical framework to the analysis of
projected dynamics in a strongly-correlated stochastic many-body
system, namely to tagged particle dynamics in a single file confined to a
tilted box.

\section{Single File Diffusion in a Tilted Box}
We now apply the theory developed in the previous section (here we use
the backward approach) to the
paradigmatic single file diffusion in a unit interval but here with a twist, namely, the
diffusing particles experience a constant
force. In particular, the full state-space is spanned by the positions
of all $N$-particles defining the state vector
$\bx_0=(x_{0,1},\ldots,x_{0,N})^T\in [0,1]^N$ and diffusion coefficients of all
particles are assumed to be equal and the thermal (white)
fluctuations due to the bath are assumed to be independent,
i.e. $\mathbf{D}=D\mathbf{1}$. In addition to being confined in a unit
interval, all particles experience the same constant force
$\mathbf{F}(\bx_0)=-\beta D F$ with $\beta=(k_BT)^{-1}$ is
the inverse thermal energy. The evolution of the Green's function is governed by the Fokker-Planck equation
\eql\ref{FPE}) equipped with the external and internal
(i.e. non-crossing) reflecting boundary conditions for the backward 
generator $\LLB=\sum_{i=1}^ND(\partial^2_{x_{0,i}}-\beta F\partial_{x_{0,i}})$:
\begin{equation}
\partial_{x_{0,1}}G(\bx,t|\bx_0)|_{x_{0,1}=0}=\partial_{x_{0,N}}G(\bx,t|\bx_0)|_{x_{0,N}=1}=0 ,\quad \lim_{x_{0,i}\to x_{0,j}}(\partial_{x_{0,i+1}}-\partial_{x_{0,i}})G(\bx,t|\bx_0)=0,
\label{BC}
\end{equation}
where we adopted the notation in \eql\ref{SGreen}). The boundary
conditions in \eql\ref{BC}) restrict the domain to a hypercone
$\bx_0\in\Xi$ such that $x_{0,i}\le x_{0,i+1}$ for $i=1,\ldots,N-1$.
The dynamics is
reversible, hence the steady state current vanishes and all
eigenvalues and eigenfunctions are real. Moreover, for
systems obeying detailed balance $\varphi(\bx)$ corresponds to the
density of the Boltzmann-Gibbs measure and it is known that
$\leftKk\equiv\ee^{-\varphi(\bx)}\rightKk$. The single file backward generator
already has a separated form
$\LLB=\sum_{i=1}^N\mathcal{L}^{\dagger}_i$ and the coupling between
particles enters solely through the non-crossing boundary condition
\eql\ref{BC}) and is hence Bethe-integrable \cite{BetheA}. However,
because the projected and latent degrees of freedom are coupled
through the boundary conditions \eql\ref{BC}) the tagged particle
dynamics is not of renewal type. 

\subsection{Diagonalization of the Generator with the Coordinate Bethe Ansatz}
Specifically, the backward generator $\LLB$ can be diagonalized
exactly using the coordinate Bethe ansatz (see e.g. \cite{Lapolla18}). To that end we first
require the solution of the separated (i.e. single particle) eigenvalue problem
$\mathcal{L}^{\dagger}_i\leftKki=-\lambda_{k_i}\leftKki$ under the
imposed external boundary conditions. Since
$\varphi(x_{0,i})=F x_{0,i}+const$ we find that
$p_{\text{ss}}(x_{0,i})=\beta F \ee^{-\beta F x_{0,i}}(1-\ee^{-\beta
  F})^{-1}$  and because of the confinement we also have
$\lambda_{0,i}=0$ 
as well as $\psi^L_{0_i}(x_{0,i})\equiv\langle
x_{0,i}|\psi^L_{0_i}\rangle=1$ and $\psi^R_{0_i}(x_{0,i})\equiv\langle
\psi^R_{0_i}|x_{0,i}\rangle=p_{\text{ss}}(x_{0,i})$. We are here
interested in the role of particle number $N$ and not of the magnitude
of the force $F$, therefore we will henceforth set, for the sake of simplicity,
$\beta F=D=1$. 
The excited
separated eigenvalues and eigenfunctions then read
\begin{equation}
\lambda_{k_i}=\pi^2k_i^2+\frac{1}{4}, \quad
\psi^L_{k_i}(x_{0,i})=\frac{\ee^{x_{0,i}/2}}{(2\pi^2k_i^2+1/2)^{1/2}}\left(\sin(k_i\pi
x_{0,i})-2k_i\pi\cos(k_i\pi
x_{0,i})\right),\quad \forall k_i\in\mathbb{Z}^+,
  \label{evals}
\end{equation}
with $\psi^R_{k_i}(x_{0,i})=\ee^{-x_{0,i}}\psi^L_{k_i}(x_{0,i})$. It
is straightforward to check that $\langle
\psi^R_{k_i}|\psi^L_{l_i}\rangle=\delta_{k_i,l_i}$. Denoting by
$\mathbf{k}=(k_i,k_2,\ldots,k_N)$ the $N$-tuple of all single-state indices $k_i$ one can show by
direct substitution that the many-body eigenvalues are given by
$\lambda_\mathbf{k}=\sum_{i=1}^N\lambda_{k_i}$ and the corresponding orthonormal many-body eigenfunctions that obey the
non-crossing internal boundary
conditions \eql\ref{BC}) have the form
\begin{eqnarray}
\boldsymbol{\psi}^L_{\mathbf{0}}(\bx_0)&=&1, \quad
\boldsymbol{\psi}^R_{\mathbf{0}}(\bx_0)=N!\prod_{i=1}^N\frac{\ee^{-x_{0,i}}}{1-\ee^{-1}}\nonumber\\
\boldsymbol{\psi}^L_{\mathbf{k}}(\bx_0)&=&\sum_{\{k_i\}}\prod_{i=1}^N\psi^L_{k_i}(x_{0,i}), \qquad  \boldsymbol{\psi}^R_{\mathbf{k}}(\bx_0)=\mathbf{m}_{\mathbf{k}}!\sum_{\{k_i\}}\prod_{i=1}^N\psi^R_{k_i}(x_{0,i}),
\label{Bethe}
\end{eqnarray}
where $\sum_{\{k_i\}}$ denotes the sum over all permutations of the
elements of the $N$-tuple $\mathbf{k}$ and $\mathbf{m}_\mathbf{k}!=\prod_i m_{k_i}!$
is the respective multiplicity of the eigenstate with $m_{k_i}$
corresponding to the number of times a particular value of $k_i$
appears in the tuple. It can be checked by explicit computation that the eigenfunctions defined in \eql\ref{Bethe})
form a complete bi-orthonormal set, that is $\langle
\psi^R_{\mathbf{k}}|\psi^L_{\mathbf{l}}\rangle=\delta_{\mathbf{k},\mathbf{l}}$
and
$\sum_\mathbf{k}\boldsymbol{\psi}^L_{\mathbf{k}}(\bx_0)\boldsymbol{\psi}^R_{\mathbf{k}}(\bx)=\delta(\bx-\bx_0)$.  

\subsection{Projection-Induced non-Markovian Tagged Particle Dynamics}
In the case of single file dynamics the physically motivated
projection corresponds to the dynamics of a
tagged particle upon integrating out the dynamics of the remaining particles. As before, we assume that the full system is prepared
in a steady state. The projection operator for the dynamics of
the $j$-th particle is therefore defined as
\begin{equation}
  \mcp_{\bx}(\delta;q_j)=\int_{\Xi}d\bx\delta(x_j-q_j)=\left[\hat{O}\prod_{i=1}^N\int_0^1dx_i\right]\delta(x_j-q_j),
  \label{taggedproj}
\end{equation}
  where the operator $\hat{O}$ orders the integration limits
  $\int_0^1dx_N\int_0^{x_N}dx_{N-1}\cdots\int_0^{x_2}dx_1$ since the
  domain $\Xi$ is a hypercone. Here, the projection is from
  $\mathbb{R}^N$ to $\mathbb{R}$.
  Integrals of this kind are easily
  solvable with the so-called 'extended phase-space integration' \cite{single,Tobias}.
  The non-Markovian Green's function is
  defined as 
\begin{equation}
Q(q_j,t|q_{0,j})=\frac{\mcp_{\bx}(\delta;q_j)\mcp_{\bx_0}(\delta;q_{0,j})G(\bx,t|\bx_0)p_{\text{ss}}(\bx_0)}{\mcp_{\bx_0}(\delta;q_{0,j})p_{\text{ss}}(\bx_0)}
\label{taggedGreen}
\end{equation}
and can be computed exactly according to \eql\ref{element}) to give
\begin{equation}
Q(q_j,t|q_{0,j})=\Psi_{\mathbf{00}}(q_{0,j})^{-1}\sum_{\mathbf{k}}\Psi_{\mathbf{0k}}(q_j)\Psi_{\mathbf{k0}}(q_{0,j})\ee^{-\lambda_{\mathbf{k}}t},
\label{StaggedGreen}
\end{equation}
where the sum is over all Bethe eigenstates and where, introducing the number of left and right neighbors, $N_L=(N-j+1)$ and $N_R=j-1$ respectively, all terms can be made explicit and read
\begin{eqnarray}
  \Psi_{\mathbf{00}}(q_j)&=&\frac{N!}{N_L!N_R!(1-\ee^{-1})^N}\ee^{-q_j}(1-\ee^{-q_j})^{N_L}(\ee^{-q_j}-\ee^{-1})^{N_R}\nonumber\\
  \Psi_{\mathbf{k0}}(q_j)&=&\frac{N!}{N_L!N_R!(1-\ee^{-1})^N}\sum_{\{k_i\}}T(q_j)\prod_{i=1}^{j-1}L(q_j)\prod_{i=j+1}^NR(q_j)
\label{all}  
\end{eqnarray}
and
$\Psi_{\mathbf{0k}}(q_j)\equiv\Psi^{\dagger}_{\mathbf{k}0}(q_j)=\frac{\mathbf{m}_{\mathbf{k}}!}{N!}\Psi_{\mathbf{k}0}(q_j)$. In
\eql\ref{all}) we have introduced the auxiliary functions
\begin{eqnarray}
T(q_j)&=&\delta_{\lambda_j,0}\ee^{-q_j}+(1-\delta_{\lambda_j,0})\frac{\ee^{-q_j/2}}{\sqrt{2}\pi\lambda_j}\left(\sin(\lambda_j\pi
q_j)-2\lambda_j\pi\cos(\lambda_j \pi q_j)\right)\nonumber\\
L(q_j)&=&\delta_{\lambda_j,0}(1-\ee^{-q_j})-2(1-\delta_{\lambda_j,0})\frac{\ee^{-q_j/2}\sin(\lambda_j\pi
   q_j)}{\sqrt{2}\pi\lambda_j}\nonumber\\
R(q_j)&=&\delta_{\lambda_j,0}(\ee^{-q_j}-\ee^{-1})+2(1-\delta_{\lambda_j,0})\frac{\ee^{-q_j/2}\sin(\lambda_j\pi
   q_j)}{\sqrt{2}\pi\lambda_j} 
\label{aux}  
\end{eqnarray}
To the best of our knowledge, equations (\ref{StaggedGreen}) to
(\ref{aux}) delivering the exact non-Markovian Green's function for the dynamics of the $j$-th particle
in a tilted single file of $N$ particles, have not yet been derived
before. Note that one can also show that
$\int_0^1dq_{j}\Psi_{\mathbf{0k}}(q_j)\Psi_{\mathbf{l0}}(q_j)\ne 0$
and hence the Chapman-Kolmogorov equation is violated in agreement
with \eql\ref{fail}) confirming that the tagged particle diffusion is indeed
non-Markovian on time-scales $t\lesssim\lambda_{\mathbf{1}}^{-1}$.

\subsection{Markovian Approximation and Degree of Broken Markovianity}
Since the projection leaves the coordinates untransformed the
effective Markovian approximation in \eql\ref{FPEp}) is particularly
simple and corresponds to diffusion in the presence of an effective
force deriving from the free energy of the tagged particle upon
integrating out all the remaining particles assumed to be in
equilibrium $\langle F(q_j) \rangle_{\bx''}=-\langle\beta D
F\delta(x_j-q_j)\rangle_{\bx''}$ or, since $-\beta D
Fp_{\text{ss}}(\bx)=\partial_{x_j}p_{\text{ss}}(\bx)$, explicitly
defined as
\begin{equation}
  \langle F(q_j) \rangle_{\bx''}=\frac{\int_{\Xi}d\bx\delta(x_j-q_j)\partial_{x_j}p_{\text{ss}}(\bx)}{\int_{\Xi}d\bx\delta(x_j-q_j)p_{\text{ss}}(\bx)}\equiv\frac{\partial_{q_j}\int_{\Xi}d\bx\delta(x_j-q_j)p_{\text{ss}}(\bx)}{\int_{\Xi}d\bx\delta(x_j-q_j)p_{\text{ss}}(\bx)}.
\label{averaging}
\end{equation}
Upon taking as before $D=\beta F=1$, and noticing that $\Psi_{\mathbf{00}}(q_j)=\int_{\Xi}d\bx\delta(x_j-q_j)p_{\text{ss}}(\bx)$ we find
\begin{equation}
\langle \LL \rangle_{\bq''}=\partial^2_{q_j}+\partial_{q_j}
\left\{\partial_{q_j}\ln\Psi_{\mathbf{00}}(q_j)\right\},\quad \langle \LLB \rangle_{\bq''}=\partial^2_{q_j}-\left\{\partial_{q_j}\ln\Psi_{\mathbf{00}}(q_j)\right\}\partial_{q_j}
\label{Markov}
\end{equation}
where the curly bracket $\{\cdot\}$ denotes that the operator inside
the bracket only acts within the bracket. The Markovian approximation of the Green's
function thus becomes $Q_{M}(q_j,t|q_{0,j})=\langle
q_{0,j}|\ee^{\langle\LLB\rangle_{\bx''}t}|q_{j}\rangle$ and is to be compared
to the exact non-Markovian Green's function (\ref{StaggedGreen}) via
the Kullback-Leibler divergence in Eq.~(\ref{KL}).

Our focus here is
to asses how the 'degree' of the projection, i.e. $d=N$, $a=1$ and
thus $d-a=N-1$ -- the number of latent
degrees of freedom (here positions of non-tagged particles) being
integrated out affects the time-dependence of the Kullback-Leibler
divergence. Since the Markovian generator cannot be diagonalized
analytically we used a finite element numerical method cross-checked
with Brownian dynamics simulations to calculate $Q_{M}(q_j,t|q_{0,j})$. The corresponding
Kullback-Leibler divergence (\ref{KL}) was in turn calculated by means of a
numerical integration. We present results for the time dependence
$\mathcal{D}_t(Q||Q_M)$ in two different representations, the
absolute (dimensionless) time $t$ and in units of the average number of
collisions $\tilde{t}=t/N^2$, tagging the third particle ($j=3$). The reason to adopt
this second choice as the natural physical time-scale is that
collisions in fact establish the effective dynamics and hence a
typical collision time sets the natural time-scale.   
\begin{figure}[!!ht]
\begin{center}  
\includegraphics[width=14.5cm]{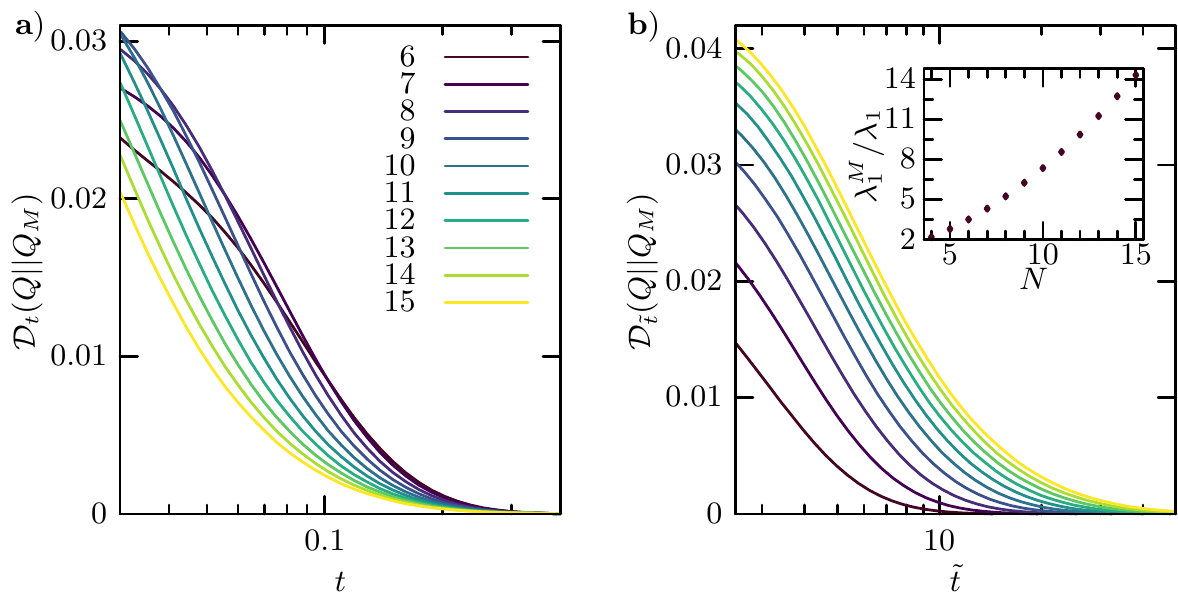}
\caption{The Kullback-Leibler divergence between the exact
  non-Markovian Green's function $Q(q_3,t|q_{0,3})$ and the Markovian
  approximation $Q_M(q_3,t|q_{0,3})$ as a function of time
  (measured in units of collision time) for increasing values of
  particle numbers $N$: a) results shown on the absolute
  (dimensionless) time-scale and b) on the natural time-scale, that is,
  expressed in units of collision time $\tilde{t}$. Inset:
  $\lambda_1^M$, the slowest
  relaxation rate of $Q_M(q_3,t|q_{0,3})$ compared to the corresponding
  eigenvalue $\lambda_1$ of the exact $Q(q_3,t|q_{0,3})$.}
\label{fig:1}
\end{center}
\end{figure}

The results
$\mathcal{D}_t(Q||Q_M)$ are shown in Fig.~\ref{fig:1}.
From Fig.~\ref{fig:1} we confirm that the Markovianity is broken
transiently (on time-scales $t\lesssim\lambda_\mathbf{1}^{-1}$, which
holds for any ergodic dynamics in the sense of generating an invariant
measure. Notably, the relaxation time $\lambda_\mathbf{1}$ does
\emph{not} depend on $N$ and is hence equal for all cases considered
here. Moreover, as expected, the magnitude of broken Markovianity
increases with the 'degree' of the projection (here with the particle
number $N$), as is best seen on a natural time-scale (see
Fig.~\ref{fig:1}b). Conversely, on the absolute time-scale the
relaxation rate of the Markovian approximation, describing
diffusion on a free energy landscape $f(q_3)=-\beta\ln\Psi_{\mathbf{00}}(q_3)$,
which can be defined as
\begin{equation}
\lambda^M_1=-\lim_{t\to\infty}t^{-1}\ln (Q_M(q_j,t|q_{0,j})-\Psi_{\mathbf{00}}(q_j))
  \label{Marrel}
\end{equation}
increases with increasing $N$ (see inset in Fig.~\ref{fig:1}b). Therefore, while both have by
construction the same invariant measure, the Markovian approximation
overestimates the rate of relaxation. This highlights the pitfall in
using free energy landscape ideas in absence of a time-scale
separation.     

\subsection{Tagged Particle Local Times Probing the Origin of Broken
  Markovianity}
In order to gain deeper insight into the origin and physical meaning
of memory emerging from integrating out latent degrees of freedom we
inspect how a given tagged particle explores the configuration
space starting from a stationary (equilibrium) initial condition. To that end we first compute the variance of local time of a
tagged particle, $\theta_t(q_j)$ in \eql\ref{localt}), given in the
general form in \eql\ref{varcor}), which applied to tagged particle
diffusion in a tilted single file reads: 
\begin{equation}
\sigma^2_t(q_j)=2\sum_\mathbf{k}\frac{\Psi_{\mathbf{0k}}(q_j)\Psi_{\mathbf{k0}}(q_j)}{\lambda_\mathbf{k}t}\left(1-\frac{1-\ee^{-\lambda_\mathbf{k}t}}{\lambda_\mathbf{k}t}\right)  
  \label{taggedlocal}
\end{equation}
where $\Psi_{\mathbf{k0}}(q_j)$ is given by \eql\ref{all}) and
$\Psi_{\mathbf{0k}}(q_j)=\frac{\mathbf{m}_{\mathbf{k}}!}{N!}\Psi_{\mathbf{k0}}(q_j)$. Note
that since the process in ergodic we have $\langle
\theta_t(q_j)\rangle=\Psi_{\mathbf{00}}(q_j)$, and because the
projected dynamics becomes asymptotically Gaussian (i.e. the
correlations between $\theta_t(q_j)$ at different $t$ gradually decorrelate) we
also have the large deviation
$\lim_{t\to\infty}t\sigma^2_t(q_j)=2\sum_\mathbf{k}\lambda_\mathbf{k}^{-1}\Psi_{\mathbf{0k}}(q_j)\Psi_{\mathbf{k0}}(q_j)\ne
f(t)$. Moreover, because of detailed balance the large deviation
principle represents an upper bound to fluctuations of time-average
observables $\sigma^2_t(q_j)\le
2\sum_\mathbf{k}\frac{\Psi_{\mathbf{0k}}(q_j)\Psi_{\mathbf{k0}}(q_j)}{\lambda_\mathbf{k}t},
\forall t$.

\begin{figure}[!!ht]
\begin{center}  
\includegraphics[width=10.5cm]{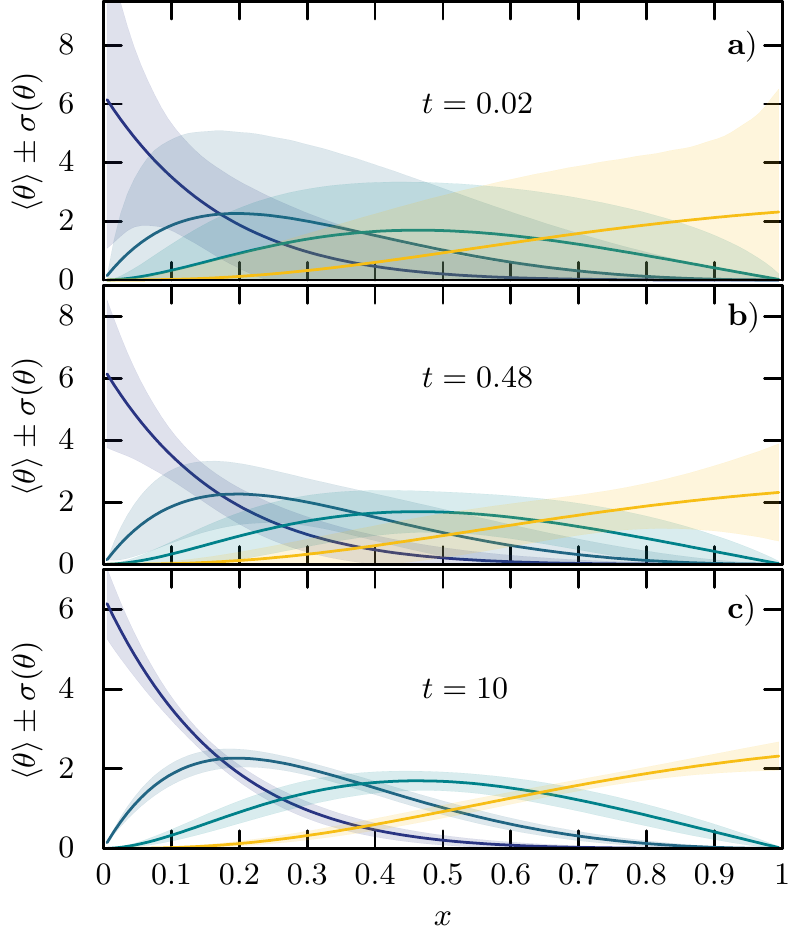}
\caption{Statistics of tagged particle local time for all members of a
single file of four particles starting from stationary initial conditions; $\langle\theta_t(q_j)\rangle$ is
represented by full lines and the region bounded
by the standard deviation $\pm\sigma_t(q_j)$ with the corresponding
shaded area. The color code is: $j=1$ violet, $j=2$ blue, $j=3$, green
and $j=4$ yellow. The relaxation time corresponds to
$\lambda_{\mathbf{1}}^{-1}\simeq 0.1$. Therefore, panel a) depicts
fluctuations on a time scale much shorter that
$\lambda_{\mathbf{1}}^{-1}$, whereas b) and c) already belong deeply
into the ergodic large deviation regime.}
\label{fig:2}
\end{center}
\end{figure}

In order to gain more intuition we inspect the statistics of $\theta_t(q_j)$ for a
single file of four particles (see Fig.~\ref{fig:2}) at different lengths of
trajectory $t$ (plotted here on the absolute time-scale). In Fig.~\ref{fig:2} we show
$\langle\theta_t(q_j)\rangle$ with full lines, and the region bounded
by the standard deviation $\pm\sigma_t(q_j)$ with the shaded area.  
The scatter of $\theta_t(q_j)$ is largest near the respective free
energy minima.

To understand further how this coupling to non-relaxed latent degrees
of freedom arises we inspect the correlations between tagged particle
histories
\begin{equation}
 \mathcal{C}_t(q_i;q_j)= \sum_\mathbf{k}\frac{\Psi_{\mathbf{0k}}(q_i)\Psi_{\mathbf{k0}}(q_j)+\Psi_{\mathbf{0k}}(q_j)\Psi_{\mathbf{k0}}(q_i)}{\lambda_kt}\left(1-\frac{1-\ee^{-\lambda_kt}}{\lambda_kt}\right),
  \label{tcor}
\end{equation}
where as before $\lim_{t\to\infty}t\mathcal{C}_t(q_i;q_j)\equiv
\overline{\mathcal{C}}_t(q_i;q_j) =\sum_\mathbf{k}\lambda_\mathbf{k}^{-1}(\Psi_{\mathbf{0k}}(q_i)\Psi_{\mathbf{k0}}(q_j)+\Psi_{\mathbf{0k}}(q_j)\Psi_{\mathbf{k0}}(q_i))\ne f(t)$ as
a manifestation of the central limit theorem, since $\theta_t(q_i)$ and
$\theta_t(q_j)$ asymptotically decorrelate. In other words, taking
$\mathcal{C}_t(q_i;q_i)\equiv\sigma^2_t(q_i)$, the complete large
deviation statistics of $\theta_t(q_i)$ (i.e. on ergodically long
time-scales) is a $N$-dimensional Gaussian with covariance matrix $t^{-1}\overline{\mathcal{C}}_t(q_i;q_j)$.

\begin{figure}[!!ht]
\begin{center}  
\includegraphics[width=15cm]{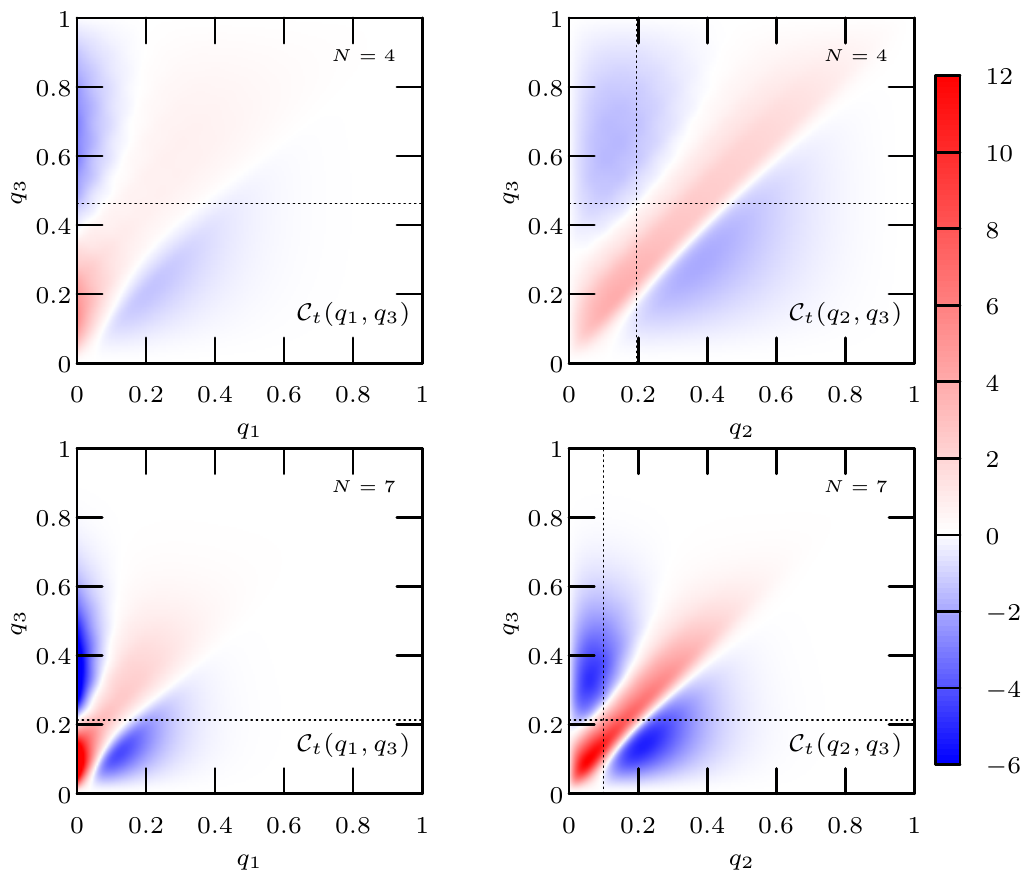}
\caption{Two-tag local time correlations $\mathcal{C}_t(q_1;q_3)$
  (left) and $\mathcal{C}_t(q_2;q_3)$ (right) for a single file of
  $N=4$ (top) and $N=7$ (bottom) particles for a (very short) trajectory length
  $t=0.01$. The relaxation time corresponds to
$\lambda_{\mathbf{1}}^{-1}\simeq 0.1$. The dashed lines denote the
  positions of the two free energy minima.}
\label{fig:3}
\end{center}
\end{figure}

\begin{figure}[!!ht]
\begin{center}  
\includegraphics[width=15cm]{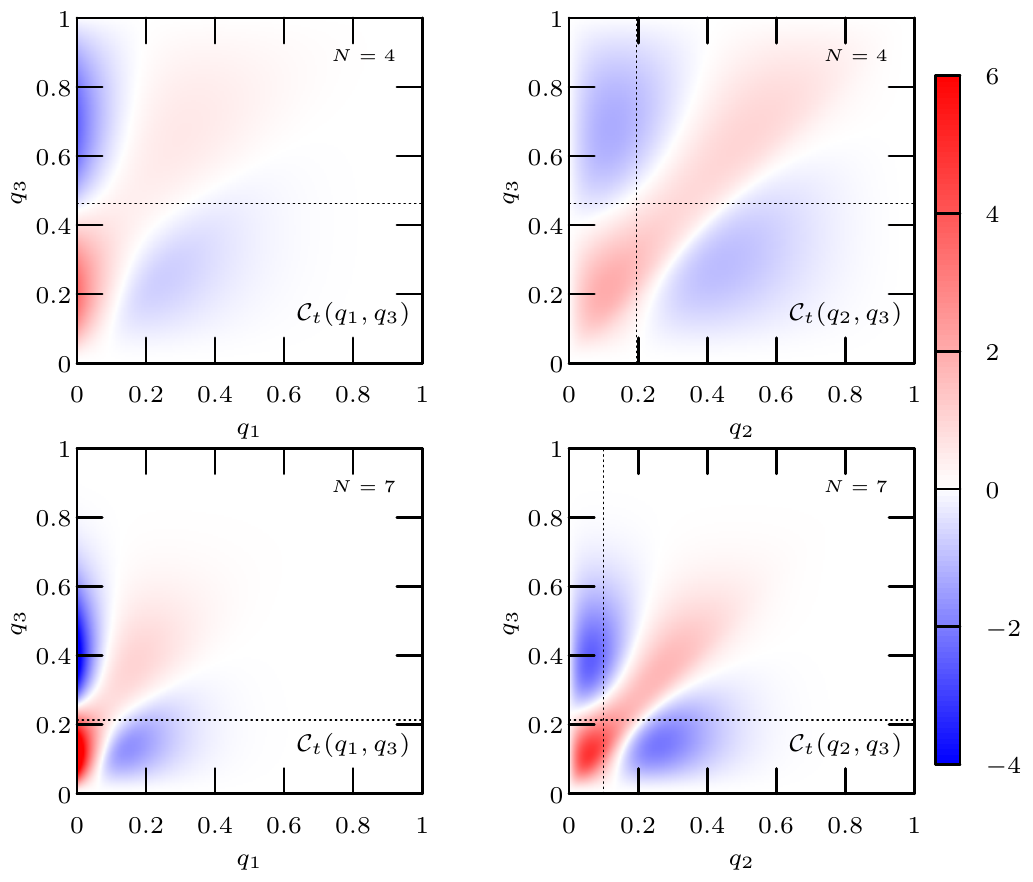}
\caption{Two-tag local time correlations $\mathcal{C}_t(q_1;q_3)$
  (left) and $\mathcal{C}_t(q_2;q_3)$ (right) for a single file of
  $N=4$ (top) and $N=7$ (bottom) particles for a trajectory length
  comparable to the relaxation time
  $t=0.12\simeq\lambda_{\mathbf{1}}^{-1}$. The relaxation time corresponds to
$\lambda_{\mathbf{1}}^{-1}\simeq 0.1$. The dashed lines denote the
  positions of the two free energy minima.}
\label{fig:4}
\end{center}
\end{figure}

To visualize these results we present in Figs.~\ref{fig:3} and
\ref{fig:4} two-tag nearest neighbor and next-nearest correlations,
$\mathcal{C}_t(q_1;q_3)$ and as $\mathcal{C}_t(q_2;q_3)$ respectively, for a single file of $N=4$ and
    $N=7$ particles at two different trajectory lengths.
We find that, alongside the fact that correlations intuitively increase
with the $N$, both the magnitude and the sign of $\mathcal{C}_t$ depend
on which particles we tag and even more so, where we tag these
particles. Along the (upward shifted) diagonal $\mathcal{C}_t$ is
positive, implying the two tagged particles along a stochastic many-body trajectory effectively (in the sense
of the local time) move together, such that if one particle spends
more time in a given region, so will the other. At fixed $F$ (here
assumed to be equal to 1) the magnitude of the
upward shift depends on which particles we tag as well as on $N$. This intuitive idea is
backed up mathematically by realizing that the lowest excited
Bethe-eigenfunctions correspond to collective ('in phase') motion
(see Eqs.~(\ref{evals}) and (\ref{Bethe})). Furthermore, defining
the free energy minima of the tagged particles with $q^{\text{min}}_i$
and $q^{\text{min}}_i$ (see dashed lines in Figs.~\ref{fig:3} and
\ref{fig:4}) we would expect, if the particles were to explore their
respective free energy minima, a peak localized at
$(q^{\text{min}}_i,q^{\text{min}}_i)$ (i.e. at the crossing of dashed line
in Figs.~\ref{fig:3} and \ref{fig:4}) .  
We find, however, that this is not the case, all together implying that
the \emph{tagged particles do not, along a many-body trajectory, explore
their respective free energy minima}. Instead, as mentioned above, they move
collectively close to each other. The collective dynamics is therefore
non-trivial and the tagged particle dynamics cannot be, at least for $t\lesssim\lambda_1^{-1}$
coarse grained to a Markovian diffusion on $-\beta\ln\Psi_{\mathbf{00}}(q_j)$, the free energy landscape
of the tagged particle $j$. Conversely, the fact that all correlations (positive and
negative) die our as $q_{i,j}\to 1$ is a straightforward consequence
of the tilting of the confining box. 

Focusing now on the dependence on the length of the trajectory
we see at very short time (much shorter than the relaxation
time) the correlations are stronger, and that positive correlations
peak further away from the two respective tagged particle free energy
minima (compare Figs.~\ref{fig:3} and \ref{fig:4}). In addition, the
maximum of $\mathcal{C}_t(q_i;q_j)$ appears to be somewhat more
localized at longer (nearly ergodic) times (see \ref{fig:4}). In
addition, the tagged particle dynamics seem to be localized more
strongly near the free energy minimum if we tag the first particle and
if $N$ is larger, presumably because of a faster relaxation due to the
presence of the wall effecting more frequent collisions with the wall,
during which the particle eventually loses memory. 

\section{Summary and Outlook}
Non-Markovian dynamics and anomalous diffusion are particularly
ubiquitous and important in
biophysical systems
\cite{Ralf1,Ralf2,Sokolov2,Klages,Godec_2014,Ralf3,Franosch,Sokolov,Ralf4,Front,Woringer,Dix,Krapf,Cox,Ilpo,Rienzo}. There,
however, it appears that the quite many non-Markovian observations
are described theoretically by phenomenological approaches with \emph{ad hoc}
memory kernels, which in
specific cases can lead to mathematically unsound or
even unphysical behavior \cite{Sokolov_master}. It therefore seems
timely and useful to provide a theoretical perspective of
non-Markovian dynamics starting from
more fine-grained principles and considering a projection to some
effective lower-dimensional configuration space.

The ideas presented
here are neither new nor completely general. Projection-operator concepts date back to the original works by Zwanzig, Mori, Nakajima,
van Kampen, H\"anggi and other pioneers. However, these seminal
contributions focused mostly on the analysis of non-Markovian evolution
equations, whereas here we provide a thorough analysis of the
manifestations of the projection on the level of Green's functions
with the aim to somewhat relieve the need for choosing a particular
model based solely on physical intuition. Furthermore, we rigorously
establish conditions under which the projected dynamics become
Markovian and renewal-type, and derive Markovian approximations to
projected generators. As a diagnostic tool we propose 
a novel framework for the
assessment of the degree of broken Markovianity as well as for the
elucidation of the origins of non-Markovian behavior.

An important remark concerns the transience of broken Markovianity,
which is a consequence of the fact that we assumed that the complete dynamics
is ergodic. First we note that (i) for any finite observation of length
$t$ it is \emph{de facto} not possible to discern whether the
observation (and the dynamics in general) will be ergodic or not on a time scale $\tau>t$. (ii)
All physical observations are (trivially) finite. (iii) In a nominally
ergodic dynamics on any finite time scale $t$, where the dynamics starting
from some non-stationary initial condition $\bx_0$ has not yet reached the
steady state (in the language of this work $t<\lambda_1^{-1}$), it is
not possible to observe the effect of a sufficiently distant confining
boundary $\partial \Omega(\bx)$ (potentially located at infinity if the
drift field $\bbf$ is sufficiently confining) that would
assure ergodicity (in the language of this work
$\forall t\ll\lambda_1^{-1}$ such that
$G(\bm{l}_{\mathrm{min}},t|\bx_0,0)\simeq 0$ where
$|\bm{l}_{\mathrm{min}}|\equiv\mathrm{min}_{\bx}|\bx_0-\partial
\Omega(\bx)|$). Therefore \emph{no generality is lost in
our work by assuming that the complete dynamics is nominally ergodic},
even in a rigorous treatment of so-called weakly non-ergodic dynamics
with diverging mean waiting times
(see e.g. \cite{Ralf1,Ralf3}) or generalized Langevin dynamics  with
diverging correlation times (see e.g. \cite{GLE2,GLE3,GLE4,GLE5,GLE6,GLE7}) on finite
time-scales. As a corollary, in the
description of such dynamics on any finite time-scale it is \emph{a
  priori} by no means necessary to assume that the dynamics is
non-ergodic or has a diverging correlation time. This does not imply,
however, that the assumption of diverging mean waiting times or
diverging correlation times cannot render the analysis of specific
models simpler.

Notably, our
work considers parent dynamics with a potentially broken time-reversal symmetry and
hence includes the description of projection-induced non-Markovian dynamics in
non-equilibrium (i.e. irreversible) systems. In the latter case the
relaxation process of the parent microscopic process might not be
monotonic (i.e. may oscillate), and it will be very interesting to
explore the manifestations and importance of these oscillations in
projected non-Markovian dynamics. 

In the context of renewal dynamics our work builds on firm mathematical foundations of Markov
processes and therefore provides mathematically and physically
consistent explicit (but notably not necessarily the most general) memory kernels derived from microscopic (or
fine-grained) principles, which can serve for the development,
assessment and fine-tuning of empirical memory
kernels that are used frequently in the theoretical modeling of
non-Markovian phenomena (e.g. power-law, exponential, stretched
exponential etc;
\cite{Ralf2,Sokolov_master}). In particular, power-law kernels are
expected to emerge as transients in cases, where the latent degrees of
freedom relax over multiple time-scales with a nearly continuous and
self-similar spectrum. Conversely, the quite strongly
restrictive conditions imposed on the microscopic (parent) dynamics that lead
to renewal dynamics, which we reveal here, suggest that renewal type
transport in continuous space (e.g. continuous-time random walks \cite{Ralf1,Ralf2}) might
not be the most abundant processes underlying projection-induced non-Markovian dynamics
in physical systems, but are more likely to arise due to some disorder
averaging. In general, it seems natural that coarse graining
involving some degree of spatial discretization should underly renewal type ideas.

From a more general perspective beyond the theory of anomalous diffusion our results are relevant for
the description and understanding of experimental
observables $a(\bq)$ coupled to projected dynamics $\bq(t)$ in
presence of slow latent degrees of freedom 
(e.g. a FRET experiment measuring the distance within a protein or a DNA
molecule \cite{Szabo}), as well as for exploring stochastic thermodynamic
properties of projected dynamics with slow hidden degrees of freedom
\cite{Seifert,Udo,Udo2}. An important field of applications of the
spectral-theoretic ideas developed here is the field of statistical kinetics in the context of first
passage concepts (e.g. \cite{Hartich_2018,Hartich_2019,Hartich_2019a}), where general results for
non-Markovian dynamics are quite sparse
\cite{HaenggiFPT,PhysRevA.38.4213,Bray,DD5,Reimann,Emerging,Satya,Guerin_2016_Mfp}
and will be the subject of our future studies.

\section*{Conflict of Interest Statement}

The authors declare that the research was conducted in the absence of any commercial or financial relationships that could be construed as a potential conflict of interest.

\section*{Author Contributions}

AL and AG conceived the research, performed the research, and wrote and reviewed the paper.

\section*{Funding}
The financial support from the German Research Foundation (DFG) through the \emph{Emmy Noether Program ‘GO 2762/1-1’} (to AG), and
an \emph{IMPRS fellowship of the Max Planck Society} (to AL) are gratefully acknowledged.

\section*{Acknowledgments}
We thank David Hartich for fruitful discussions and critical reading
of the manuscript. AG in addition thanks Ralf Metzler for introducing
him to the field on anomalous and non-Markovian dynamics and for years
of inspiring and encouraging discussions.




\bibliographystyle{frontiersinHLTH&FPHY} 

\end{document}